\documentclass{article}

\usepackage[utf8]{inputenc}
\usepackage[round]{natbib}
\usepackage{color}
\usepackage{comment}
\usepackage{graphicx}
\usepackage{float}
\usepackage[hidelinks]{hyperref}

\usepackage{authblk}

\providecommand{\keywords}[1]
{
  \textbf{Keywords: } #1
}

\title{A scientometric analysis of the effect of COVID-19 on the spread of research outputs}

\author[1, *]{Zammarchi G.}
\author[1]{Carta A.}
\author[2]{Columbu S.}
\author[1]{Frigau L.}
\author[2]{Musio M.}

\affil[1]{\small Department of Economics and Business Science, University of Cagliari, Cagliari, Viale Sant’Ignazio 17, 09123, Cagliari, Italy}
\affil[2]{Dipartimento di Matematica e Informatica, Università degli Studi di Cagliari, Via Ospedale 72, 09124, Cagliari, Italy}
\affil[*]{Corresponding author: gp.zammarchi@unica.it}

\date{}

\begin{document}

\maketitle

\begin{abstract}
\noindent The spread of the Sars-COV-2 pandemic in 2020 had a huge impact on the life course of all of us. This rapid spread has also caused an increase in the research production in topics related to COVID-19 with regard to different aspects. Italy has, unfortunately, been one of the first countries to be massively involved in the outbreak of the disease. In this paper we present an extensive scientometric analysis of the research production both at global (entire literature produced in the first 2 years after the beginning of the pandemic) and local level (COVID-19 literature produced by authors with an Italian affiliation). Our results showed that US and China are the most active countries in terms of number of publications and that the number of collaborations between institutions varies according to geographical distance. Moreover, we identified the medical-biological as the fields with the greatest growth in terms of literature production. Furthermore, we also better explored the relationship between the number of citations and variables obtained from the data set (e.g. number of authors per article). Using multiple correspondence analysis and quantile regression we shed light on the role of journal topics and impact factor, the type of article, the field of study and how these elements affect citations.
\end{abstract}

\keywords{scientometric analysis, COVID-19, MCA, quantile regression, citations}

\section{Introduction}
\noindent The beginning of the Coronavirus disease 2019 (COVID-19) pandemic was declared by the World Health Organization (WHO) on March 11, 2020. Later, new evidence showed that the first cases of human transmission of the virus occurred in late 2019 and were mainly located in Wuhan, China \citep{Guan}. To date, the number of cases reported by WHO worldwide has exceeded 600 million, with more than 6 million deaths related to the virus \citep{who}. Almost all governments in the world have taken countermeasures to face the spread of this highly infectious virus, ranging from quarantine to wearing masks or implementing social distances. The consequences of the pandemic have been faced in several aspects of our life. It is therefore not surprising that a very high number of researchers in all specialties (both medical and non-medical) started contributing to this research. In less than three years, the number of articles published on this topic has grown exponentially and researchers from all over the globe have contributed to study different aspects related to COVID-19 (which is related to the disease) or SARS-CoV-2 (which indicates the virus) \citep{Fassin2021}. 
Given the sheer number of single studies that focus on innumerable aspects related to COVID-19, we performed a scientometric analysis that gives a comprehensive picture of this relevant topic. 
Many authors have tried to give a definition of scientometric analysis and also to highlight the similitudes and differences with bibliometric or infometric analysis \citep{TagueSutcliffe, VanRaan, Hood}. In this paper, we consider a scientometric analysis as the quantitative study of the disciplines of science based on published literature and communication \citep{Thompson}. Some of the main results of a bibliometric analysis are the identification of new or emerging areas of scientific research, their development and trends over time, or the geographical and organizational distribution of research \citep{Thompson}. When a new topic of study arises, it is necessary to collect and analyze, after an adequate period of time, the literature related to that topic in order to have a general idea of how scholars from all over the world are analyzing that event or phenomenon. The main aims are to portray a comprehensive picture of the growth of the scientific literature, to illustrate the structure and the relationships between e.g. articles, authors, or institutions in a research field or in a specific topic. For all these reasons, a scientometric analysis is one of the most important approaches for evaluating the scientific production. 

Several researchers have studied the literature on COVID-19 from different points of view, using a scientometric approach. Some authors focused on the repercussions of the pandemic on a specific aspect of the society, such as the economic or the educational aspect. For example, Hashemi and colleagues \citep{Hashemi} evaluated the impact in the management field. They retrieved all articles on COVID-19 on Web of Science or Scopus and selected only those related to business, management, or accounting, showing the main themes for 2020 and 2021 on individual, organizational and societal levels. They found that in 2020 researchers were more focused on experiences and coping with COVID-19, while in 2021 studies were mostly about the acceptance of new rules in the workplace and business environment. Su and colleagues \citep{Su} analyzed articles related to the impact that COVID-19 had on financial, operational, and other aspects of enterprises’ management. Financial liquidity, market channel expansion, supply chain stability, and efficiency, were all aspects affected by COVID-19. The authors also highlighted the importance of a leadership capable of adapting to change and exploiting opportunities. Zhang et al. \citep{Zhang} identified 1,061 documents that evaluated the effects of COVID-19 on online higher education in 103 countries. The authors focused on challenges of online education (in particular for students with impairments), innovative pedagogies in online learning, and distribution of literature, suggesting that open access represents a tool to reduce barriers in spreading knowledge. The study reported that, due to the pandemic, a higher number of scholars is exploring a wide range of topics related to the changes in online higher education. Gómez-Domínguez and colleagues \citep{GomezDominguez} focused on the analysis of the scientific production on teachers’ stress during the COVID-19 outbreak. This study showed a high interest in the topic of stress and burnout, highlighting that many studies were related to mental health, coping strategies and other measures to mitigate the effects of the pandemic and improve teachers’ well-being.

Other authors narrowed the analysis geographically, for example analyzing only a specific country, a group of countries, a (broad) area, or a continent. Shamsi and colleagues \citep{Shamsi} analyzed publications from 3,450 researchers retrieved from WoS, PubMed, and Scopus. They mainly used graphical representation of networks of countries, authors, and words to describe Iranian publications and reported that, for example, compared to other countries, Iran had larger research teams. Other authors focused on Asian countries such as India, Arab Emirates, Korea or zones such as Southeast Asia. Raju and Patil \citep{Raju} reported that the most cited Indian articles in 2020 were those related to virology, diagnosis, treatment or clinical features, while general studies on epidemiology or pandemic received less citations. Al-Omari and colleagues \citep{AlOmari} described the research activity of the United Arab Emirates-affiliated researchers from 2020 to 2022. They reported that most authors affiliated with the United Arab Emirates collaborated with colleagues from the same country and that the main international collaboration were with the US and England. Kim and Jeong \citep{Kim} reported results from a bibliometric analysis conducted on Korean articles to identify the collaborations between Korean and international authors and explore clusters of institutions, journals, and topics. They found that Korean authors mainly collaborated with United States authors. Tantengco \citep{Tantengco} analyzed more than 700 articles from Southeast Asia in 2020 and reported that Malaysia was the most active country with respect to the number of publications, while Singapore affiliated authors received a higher number of citations.
Chiu and Ho \citep{Chiu} collected all studies based on Latin America published in 2020 and found that Brazil was one the most active countries, while Corrales-Reyes and colleagues \citep{CorralesReyes} focused only on Cuba and found that publications with Cuban leadership were less likely to have a major impact. Another very recent article \citep{Chatterjee} also focused on countries that are in the same geographical area. The authors described the scientific production of all the English-speaking Caribbean countries and observed that more than 50\% of the research based on Caribbean originated from Trinidad and Tobago or Jamaica.
Stojanovic \citep{Stojanovic} assessed which were the main areas of research during the first five months of 2020 in Canada. The author reported that the main topics were infection, prevention, therapeutics, among others, and that, at that time, there was a gap in the literature about diagnostic and vaccines. 
In 2021, Turatto \citep{Turatto} analyzed articles retrieved in PubMed and Scopus in the early phase of COVID-19 in Italy, focusing mainly on articles published by authors with an Italian affiliation but also on articles analyzing Italian data. The authors reported that many Italian articles were about the management aspect of the pandemic, while non-Italian articles were mostly epidemiological studies.

Few articles analyzed the global literature. Some of these publications were about the early stages of COVID-19 outbreak \citep{Aviv2021, Haghani, Furstenau, Tran}. They all highlight the fact that in a few months the number of articles published on this issue increased exponentially, with hundreds or thousands of articles published in a very short timeframe. In 2020, another bibliometric analysis by Hamidah and colleagues \citep{Hamidah} reported that, since the beginning of the pandemic, China, UK and US were among the top contributors to the COVID-19 literature, and also that the main topics were related to public health and laboratory studies. In 2021, Wang and colleagues \citep{Wang} analyzed data referred to 2020 through WoS and several preprint platforms (bioRxiv, medRxiv, Preprints, and SSRN) to show the global trends in COVID-19 research. They reported US to be the most active country (followed by China) with respect to the number of contributions to the literature. 
In our study, we followed a glocal approach. First, we presented a comprehensive and exhaustive overview of the global literature on COVID-19 and next we focused on Italy, as it was one of the first Western countries to be severely affected by COVID-19.

Aims of this study were the following: 1) to conduct an extensive and up to date scientometric analysis of studies investigating different aspects related to the COVID-19 pandemic outbreak; 2) to investigate the main topics studied in these articles; 3) to identify the most active countries and institutions, as well as their relationships; 4) to conduct a case study on articles with authors affiliated with an Italian institution based on the fact that this country was among the first to be severely affected by the COVID-19 outbreak and 5) to investigate the relationship between the number of citations and different characteristics of the selected articles.

\section{Methods}

\subsection{Data collection and processing}
\noindent We conducted a literature search on the Web of Science Core Collection (WoS) online database, updated to the 31st of May 2022, to retrieve any scientific article or review on COVID-19. We used the following search strategy: \textit{covid* OR “corona\$virus disease *19" OR sars-cov-2}. We searched for studies mentioning these terms in the Topic field (title, abstract, author keywords and keyword plus\footnote{added by WoS, not by the authors as for the Keywords}). Only scientific articles and reviews, written in English, published in 2020, 2021 or 2022 were retained. For each article, we extracted the following characteristics: title, abstract, keywords plus, authors’ affiliations, year of publication, type of publication (article or review), journal title, journal category or categories based on the classification from the Journal citation report (JCR), and the number of citations. The JCR classification includes 254 research categories, which are further assigned to 21 groups. Each journal can be assigned to one or more research categories and each research category can be part of one or more groups. For each article of our data set published in a journal included in the JCR classification, we assigned one or more of the 21 groups based on the research categories reported in the data set. In addition, we retrieved the journal impact factor based on the JCR 2021.

\subsection{Bibliometric analysis}
The bibliometric analyses were conducted with the Bibliometrix package \citep{Aria} version 3.1 in R version 4.1.2. \citep{Rcoreteam2021} and the Biblioshiny shiny app \citep{Aria}. We identified countries and institutions with the highest number of published articles and/or citations (based on the affiliation of the authors). The collaboration networks of countries and institutions were generated using the Louvain clustering algorithm implemented in Bibliometrix, setting the number of nodes to 20, for better clarity of representation. A co-occurrence network of keywords plus was also generated using the 20 most frequent words, after excluding words present in the search query.
 
Next, we conducted a case study on a subset of the data set including only articles with at least one author affiliated with an Italian institution. In order to do this, we selected only articles with at least one author for which the word Italy was reported in the address field. For these authors, we collected additional information relatively to the disciplinary scientific sector (SSD, in Italian “Settore Scientifico Disciplinare") from the CINECA official database of employees' website \citep{cineca}. This database contains information only for authors affiliated with an Italian university. For each article we computed the total number of authors as well as the number of authors with or without an Italian affiliation. In addition, we codified variables regarding first and last author. The variable is set to one if the author in first/last position has an Italian affiliation and zero otherwise.

\subsection{Multiple Correspondence Analysis}
Multiple Correspondence Analysis (MCA) is a type of factor analysis that is particularly useful for analyzing data with many categorical variables, it is similar to principal component analysis, but is specifically designed for categorical data. MCA is used to identify patterns and relationships in the data, and to reduce the dimensions of the data by projecting it onto a lower-dimensional space. The first step in conducting MCA is to create a multi-way contingency table of the categorical variables, then the table is transformed into an indicator matrix or a Burt matrix and finally, a simple Correspondence Analysis is applied to one of them \citep{benzecri1969statistical}. This method allows one to represent graphically the transformed data in a bi-plot where each point represents a category, and the position of the points reflects the relationship between the categories. This reduces the dimensionality of the data, and helps to analyze the pattern of relationships among a multitude of categorical dependent variables.

\subsection{Analysis of citations}

We conducted regressive analyses to identify variables associated with the number of citations. In order to take into account the time passed between the publication of each article and the date at which articles were retrieved, the number of citations was divided by the number of months from the publication date, this rate of citations was used as response in the models estimated. Predictors included the total number of authors, the JIF (Journal Impact Factor) based on the JCR Report 2021 as well as the JCR research group(s) of journals in which articles were published. Articles published in journals with no JIF, no listed authors, not published from at least one month or articles with missing or no citations were excluded, leading to a total of 98,054 included documents. The asymmetry of the response suggested that instead of focusing on its conditional mean, through the estimation of a standard linear regression, was more appropriate to analyse some of the quantiles of its distribution. This is done bymeans of linear quantile regression models \citep{Koenker1978, Davino2013}, which allow to estimate a different effect of the predictors for each selected quantile of the response. We estimated a separate model for the three quartiles of the rates of citations, this can give an understanding of possible different behaviours for highly, median or less cited papers. The models were fitted through the \textit{quantreg} package of \textit{R} (\cite{Koenker:2009}) for a selected subset of the whole data-set. We repeated the same approach for Italian publications.



\section{Results}

\subsection{Bibliometric analysis on the whole data set}
\begin{figure}[h]
\centering
  \includegraphics[width=0.95\textwidth]{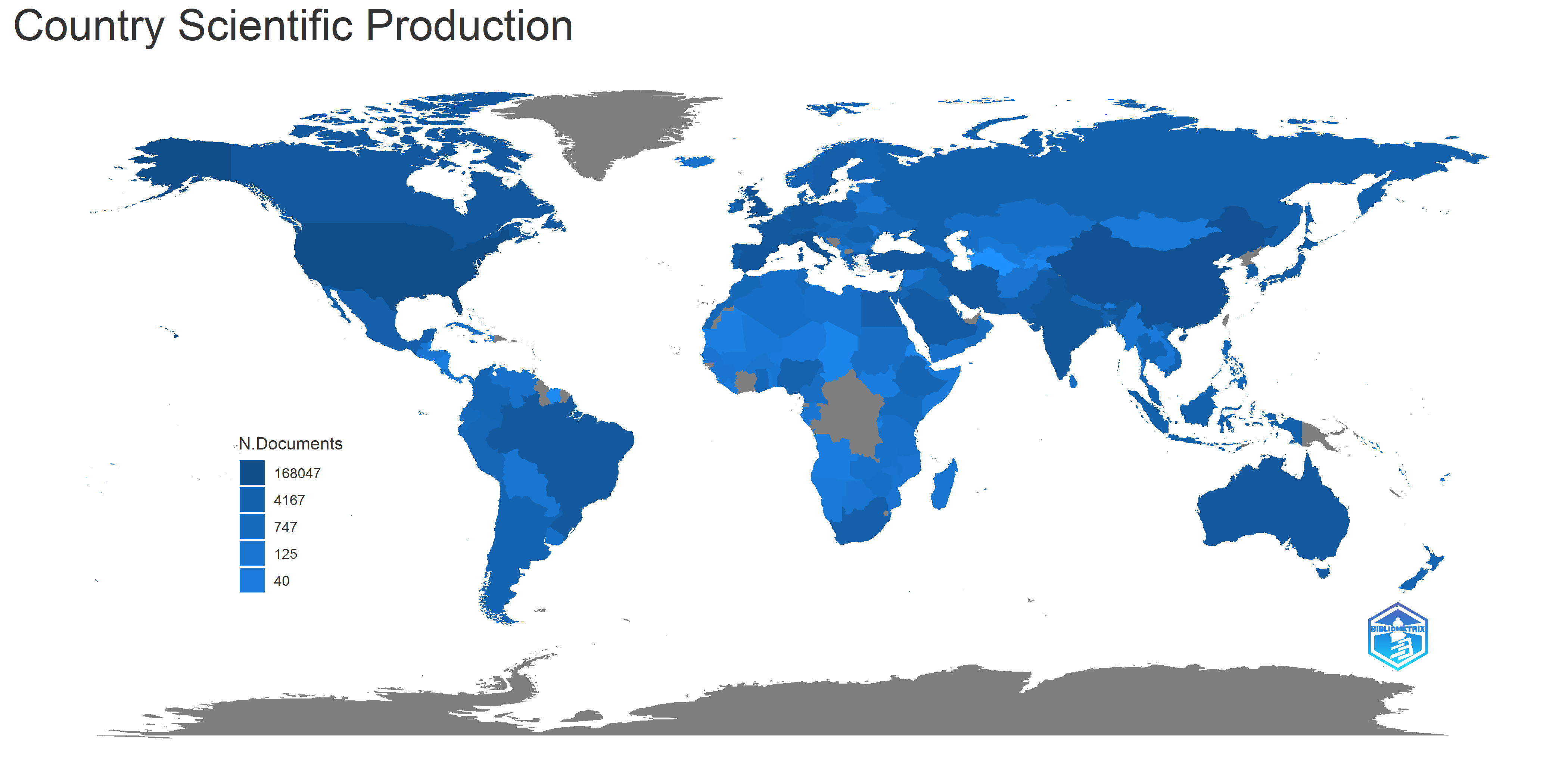}
\caption{Countries for which a higher number of documents based on the affiliation of the corresponding authors are shown in darker blue. In this representation a document might be counted multiple times, once for each author. }
\label{fig:world_nDoc}
\end{figure}

\noindent Our search retrieved 209,124 articles and reviews. After removing documents that did not satisfy our inclusion criteria (publication year equal to 2020, 2021 or 2022 and written in English), analyses were conducted on a total of 184,098 documents. Table \ref{tab:world_nDoc} shows the number of documents retrieved for each country based on the affiliation of the corresponding author. The main countries with respect to the number of articles are USA (40,060, 22\%), China (19,938, 11\%), United Kingdom (UK) (11,334, 6\%) and Italy (11,232, 6\%).
A similar scenario is represented in Figure \ref{fig:world_nDoc}, which shows the number of documents for each country based on the affiliation of any author. 

For each of the top-ten countries based on the number of retrieved documents, we computed the number of single country publications (SCP, i.e. articles with no international collaborations), the number of multiple country publications (MCP, i.e. articles with international collaborations) as well as the ratio between MCP and the total number of documents. We can see that, among the top ten countries, Australia, UK and Germany showed the highest MCP ratio (i.e. a higher percentage of documents for these countries included international collaborations). The USA were the first country based on the number of articles, but only a small percentage of them included international collaborations (17.8\% compared to e.g. 37.5\% for e.g. Australia). 

\begin{table}[]
\begin{tabular}{lrrrr}
\hline
\textbf{Country} & \textbf{Articles} & \textbf{SCP} & \textbf{MCP} & \textbf{\% of Int. Coll.} \\
\hline
USA              & 40,060             & 32,926        & 7,134         & 0.1781              \\
China            & 19,938             & 14,965        & 4,973         & 0.2494              \\
United Kingdom   & 11,334             & 7,482         & 3,852         & 0.3399              \\
Italy            & 11,232             & 8,704         & 2,528         & 0.2251              \\
India            & 10,427             & 8,679         & 1,748         & 0.1676              \\
Canada           & 5,400              & 3,623         & 1,777         & 0.3291              \\
Spain            & 5,233              & 4,067         & 1,166         & 0.2228              \\
Germany          & 5,178              & 3,425         & 1,753         & 0.3385              \\
Turkey           & 4,850              & 4,477         & 373          & 0.0769              \\
Australia        & 4,650              & 2,908         & 1,742         & 0.3746              \\
\hline
\end{tabular}
\caption{Total number of articles retrieved in the search, SCP, MCP, and percentage of articles with international collaborations. Abbreviations: SCP, Single Country Publications, MCP, Multiple Country Publications, \% of Int. Coll., Percentage of International Collaborations}
\label{tab:world_nDoc}
\end{table}

\begin{figure}[]
\centering
  \includegraphics[width=0.8\textwidth]{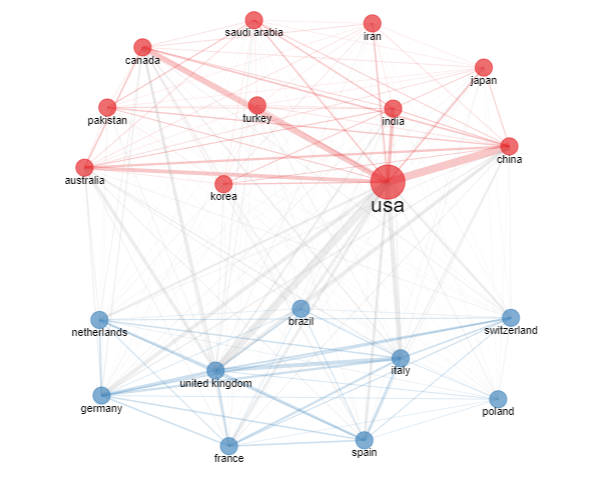}
\caption{Country collaboration network}
\label{fig:world_netCountry}
\end{figure}

\begin{figure}[]
\centering
  \includegraphics[width=0.8\textwidth]{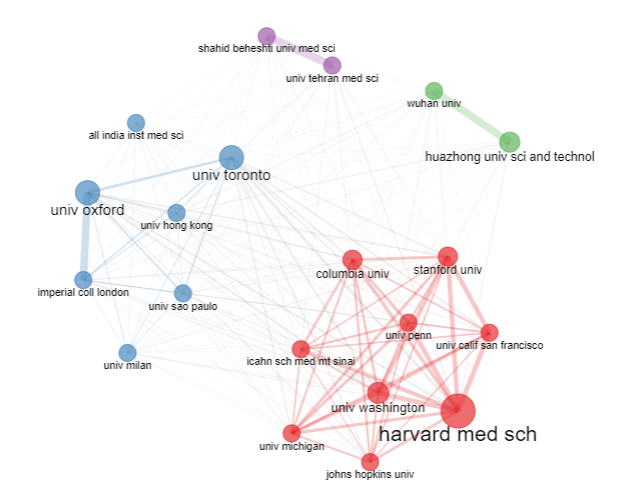}
\caption{Institution collaboration network}
\label{fig:world_netInstitution}
\end{figure}

Studies were carried on by authors from 88,011 institutions located in 165 countries. We created a country (Figure \ref{fig:world_netCountry}) and an institution (Figure \ref{fig:world_netInstitution}) collaboration network, based on the approach implemented in \citep{Aria}. In the graphical representation of these networks, each circle represents a country or an institution, the size of the circle is proportional to the number of documents and the thickness of the lines represents the strength of the relationship between two countries or institutions. With respect to the country collaboration network, we identified two main clusters shown in Figure \ref{fig:world_netCountry}. One cluster (reported in red in Figure \ref{fig:world_netCountry}) included the USA, Australia, Canada and different Asian countries (e.g., China, Japan, India and Korea). The other cluster (reported in blue in Figure \ref{fig:world_netCountry}) included UK, Brazil and different European countries (e.g., Italy, Germany, France and Spain). When constructing the institution collaboration network, we identified four clusters (Figure \ref{fig:world_netInstitution}). The cluster including the highest number of institutions (reported in red in Figure \ref{fig:world_netInstitution}) only included institutions based in the USA (e.g., Harvard, University of Washington, Stanford). The second largest cluster (reported in blue in Figure \ref{fig:world_netInstitution}) included a variety of institutions from United Kingdom (University of Oxford and Imperial College London), Canada, Italy, Brazil, India and Hong Kong. Interestingly, this cluster included institutions from several of the top ten countries based on the number of COVID-19 articles (Table \ref{tab:world_nDoc}). The two smaller clusters contained institutions from Asian countries (colored in purple and green in Figure \ref{fig:world_netInstitution}).

Table \ref{tab:world_sources} shows the 10 most relevant sources in terms of number of documents in the topic. The top three resulted to be the “International Journal of Environmental Research and Public Health", “PLOS ONE" and “Sustainability". All were open access journals. 

\begin{table}[]
\begin{tabular}{lrrrr}
\hline
\textbf{Journal} & \textbf{N. of articles} & \textbf{JIF} \\
\hline
IJERPH & 4,136 & 3.39 \\
PLOS ONE & 2,970 & 3.24 \\
Sustainability   & 2,060 & 3.25         \\ 
Scientific Reports & 1,881 &  4.38 \\
Frontiers in Psychology & 1,618 & 2.99 \\
Frontiers in Public Health & 1,494 &  3.71 \\
BMJ Open & 1,222 & 2.69 \\
Vaccines & 1,146 & NA \\
Journal of Clinical Medicine & 1,137 & 4.24 \\
Cureus Journal of Medical Science & 1,084 &  NA \\
\hline
\end{tabular}
\caption{Top ten sources based on the number of published articles. Abbreviations: IJERPH, International Journal of Environmental Research and Public Health; JIF, journal impact factor 2020}
\label{tab:world_sources}
\end{table}

\begin{figure}[H]
\centering
  \includegraphics[width=0.8\textwidth]{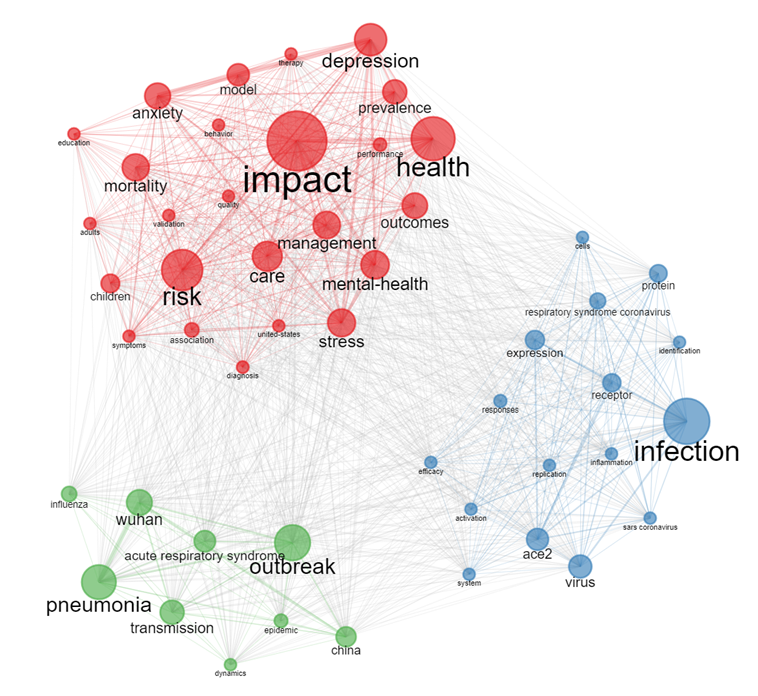}
\caption{Co-occurrence network constructed using the 50 most frequent keywords plus }
\label{fig:world_netKeywords}
\end{figure}

In the top twenty documents of the collection of COVID-19 articles, the most cited document was \citep{Zhou2020}.\\ As a last step, using the 50 most frequent keywords plus, we created a word co-occurrence network shown in Figure \ref{fig:world_netKeywords}. 

Each word is represented with a circle, and the size of the circles is proportional to the number of documents including the word. The degree of the relatedness between two words is indicated by the thickness of the line connecting the circles. As shown in Figure \ref{fig:world_netKeywords}, three main clusters of words were identified. One mostly included words related to the impact of the infection (e.g. risk, impact, care, health, mortality, outcomes, management) and in particular to the impact on mental health (depression, anxiety, stress, mental health, represented in red in Figure \ref{fig:world_netKeywords}). A second cluster included words related to epidemiological (e.g. transmission, outbreak, Wuhan, China) or clinical (pneumonia, acute respiratory syndrome) aspects and is represented in green in Figure \ref{fig:world_netKeywords}. The last cluster included words related to biological mechanisms (e.g. protein, expression, ace2, receptor, cells, replication and inflammation, represented in blue in Figure \ref{fig:world_netKeywords}).

\subsection{Case study on articles with authors affiliated with an Italian institution}
Among the 184,098 included documents, 14,916 included at least one author with an Italian affiliation.
Table \ref{tab:italy_nDoc} shows the countries for which we identified the highest number of collaborations among articles with Italian affiliated authors. The USA were the country for which the highest number of collaborations was identified, followed by UK and Spain (Table \ref{tab:italy_nDoc}). 

\begin{table}[h]
\begin{tabular}{lr}
\hline
\textbf{Country} & \textbf{N. of articles} \\
\hline
USA              & 2,251                        \\
UK            & 1,860                \\
Spain   & 1,201                  \\ 
Germany            & 1,079            \\
France            & 1,040          \\
Switzerland                  &  756                      \\
Netherlands            & 687            \\
Canada         & 633            \\
China              & 599                          \\
Australia       &  580                       \\

\hline
\end{tabular}
\caption{International collaborations between autors with Italian affiliation and authors from other countries }
\label{tab:italy_nDoc}
\end{table}

\begin{figure}[h]
\centering
  \includegraphics[width=0.8\textwidth]{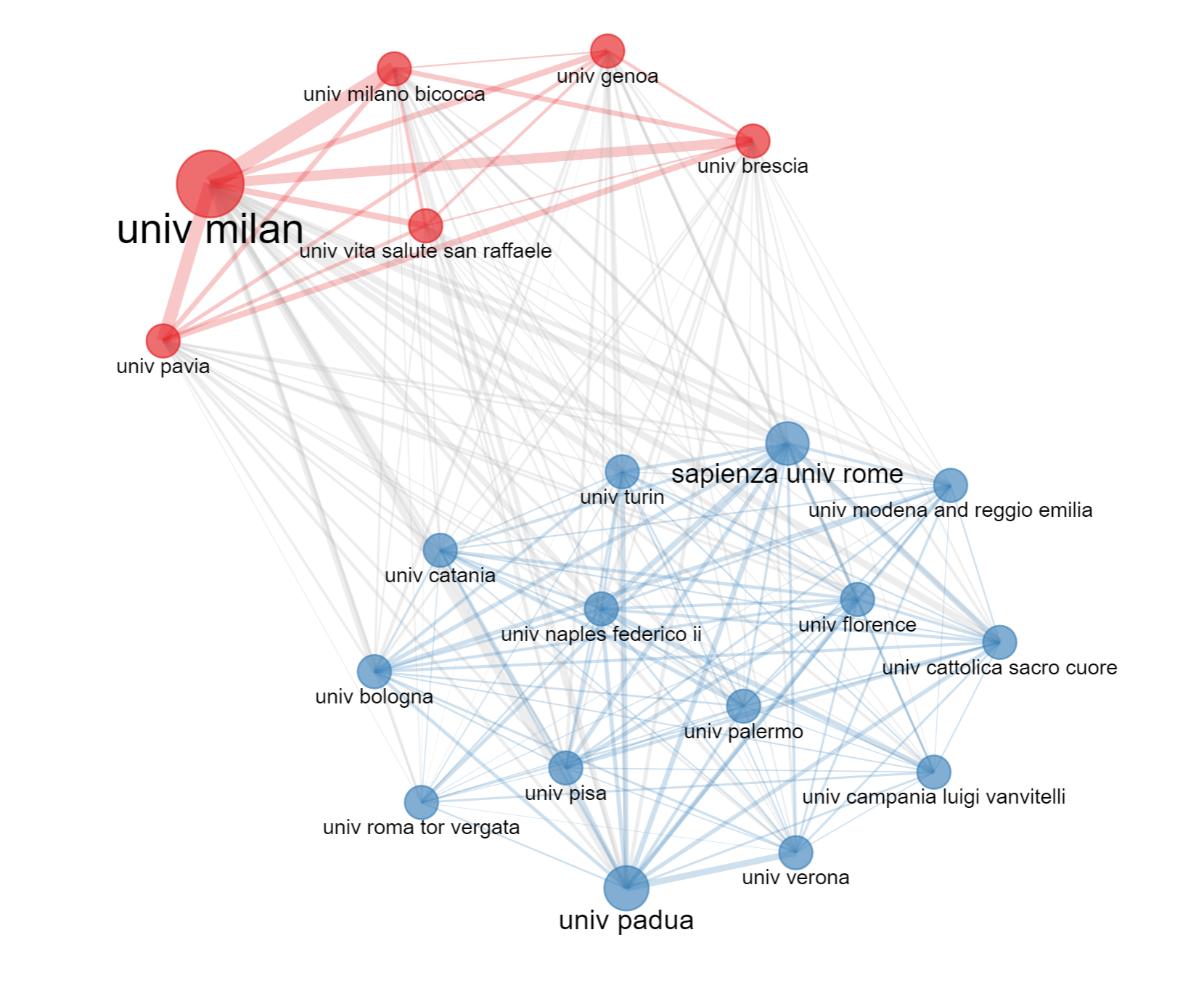}
\caption{Network of institutions among articles with authors affiliated with Italian institutions}
\label{fig:italy_netInstitution}
\end{figure}

\begin{figure}[H]
\centering
  \includegraphics[width=0.8\textwidth]{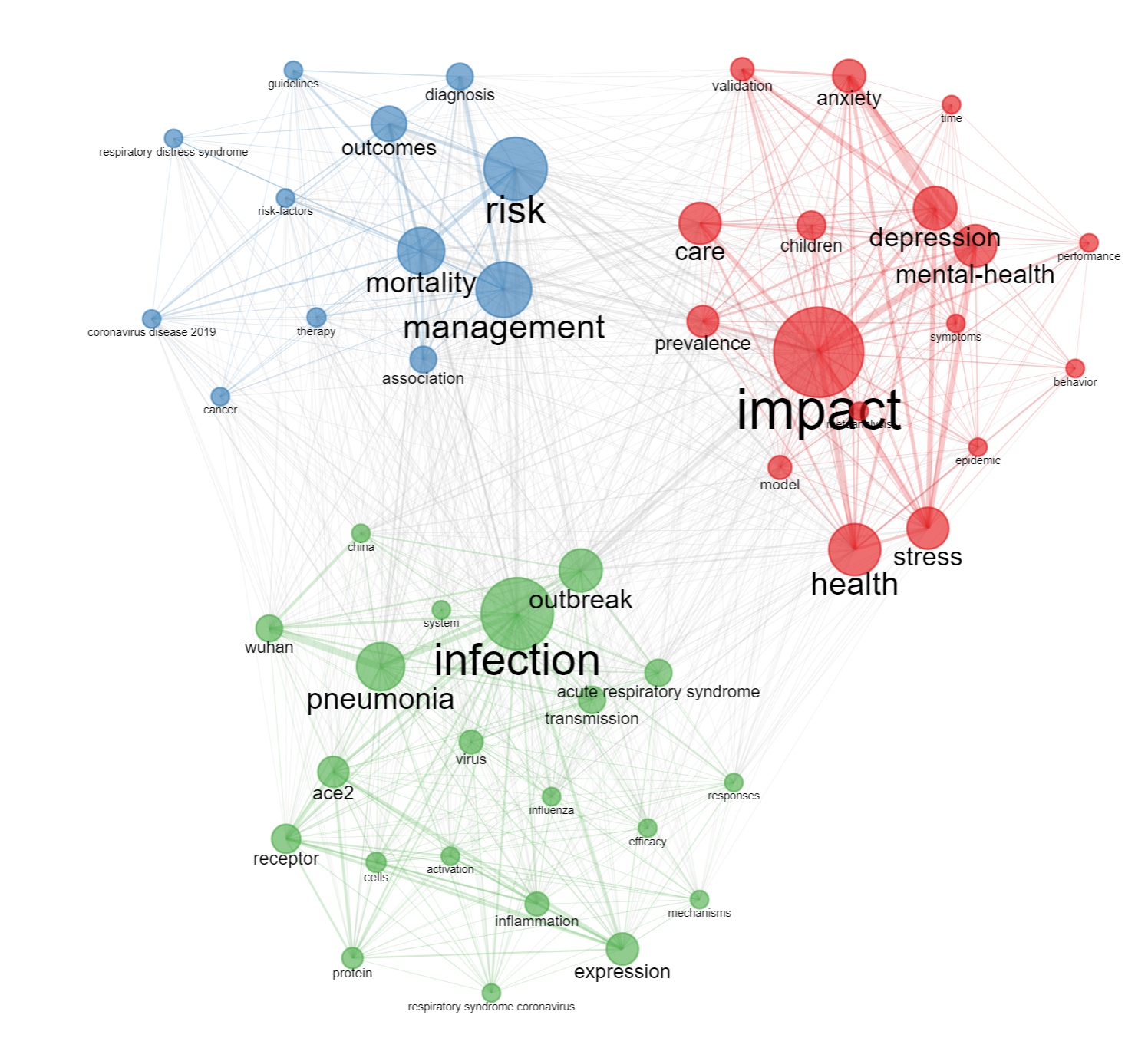}
\caption{Co-occurrence network constructed using the 50 most frequent keywords plus among articles with authors affiliated with Italian institution}
\label{fig:italy_netKeywords}
\end{figure}

Figure \ref{fig:italy_netInstitution} shows the institution collaboration network based on authors with an Italian affiliation. Two main clusters of institutions were identified: one including mostly universities from northern Italy (shown in red) and one including universities from either northern, central or southern parts of Italy (shown in blue). 
The co-occurrence network constructed using the 50 most frequent keywords plus is shown in Figure \ref{fig:italy_netKeywords}. 

The network included three clusters and was similar to the one constructed using the whole data set (Figure \ref{fig:world_netKeywords}). However, in the network constructed using the Italian subset, the two clusters previously identified as related to epidemiological aspects and biological mechanisms were merged in a single cluster (shown in green in Figure \ref{fig:italy_netKeywords}). The third cluster included words related to the management and consequences of the infection (e.g. risk, mortality, management, outcomes, diagnosis) and is shown in blue in Figure \ref{fig:italy_netKeywords}. 

Finally, for each author affiliated with an Italian university, we retrieved her/his scientific-disciplinary sector (SSD\footnote{\url{https://www.miur.gov.it/settori-concorsuali-e-settori-scientifico-disciplinari}}).

\begin{figure}[h]
\centering
  \includegraphics[width=0.8\textwidth]{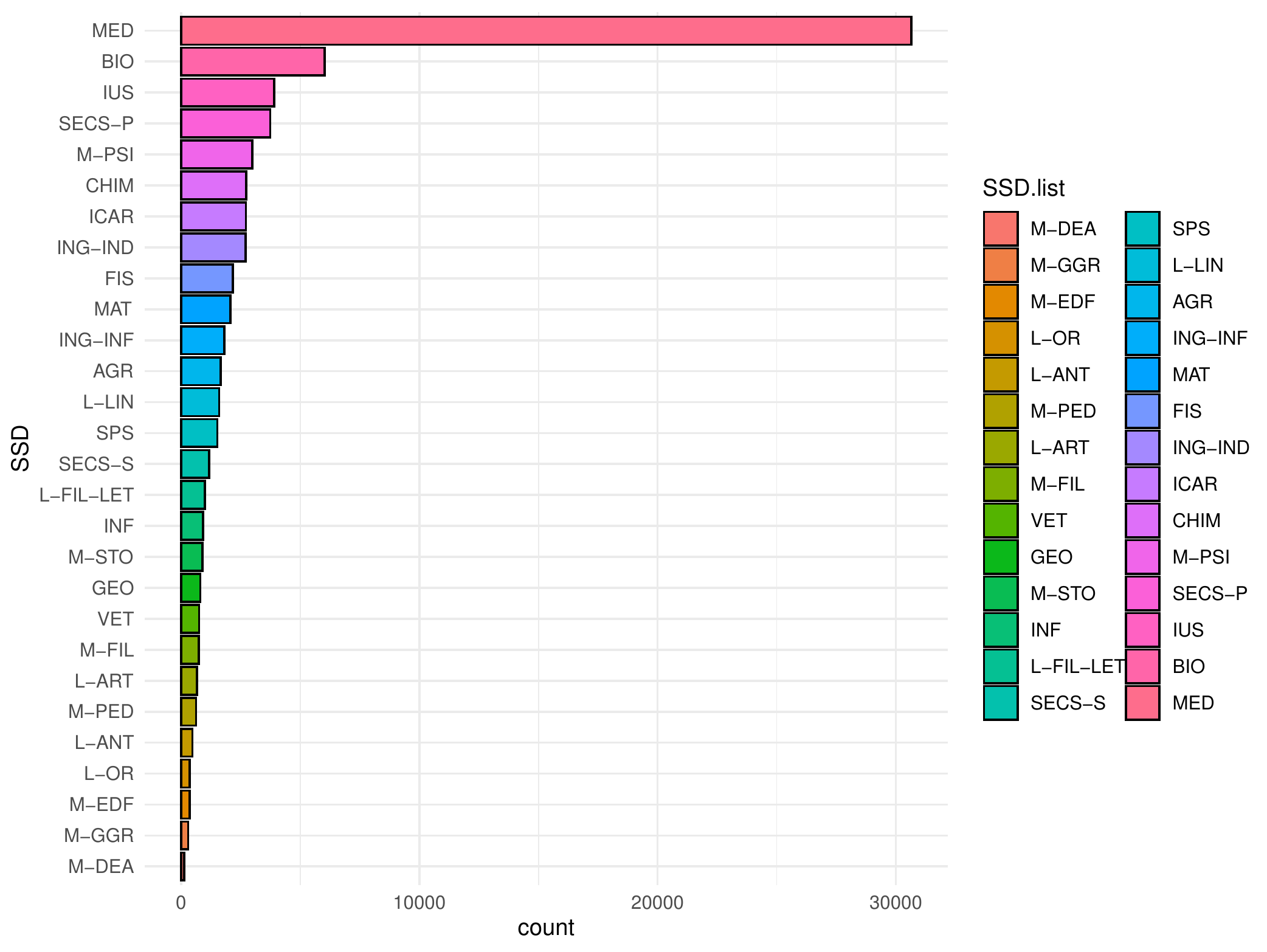}
\caption{Number of researchers in each SSD. Abbreviations: SSD, Settore Scientifico Disciplinare (disciplinary scientific sector)}
\label{fig:italy_SSD}
\end{figure}

In the Italian Higher Education system every researcher and professor is, in fact, necessarily classified in one among 383 different sectors that describe their research domain. As can be expected, the most represented sector was medicine, followed by biology and other sectors such as law and economics (Figure \ref{fig:italy_SSD}). For a complete list of the SSD with their description see the CUN (Italian National University Council) website\footnote{\url{https://www.cun.it/uploads/storico/settori\_scientifico\_disciplinari\_english.pdf}}.

\subsection{Multiple Correspondence Analysis}
In this section, by using MCA we investigate how the variables 
Citation Parameter and Topic are associated with 
Journal Impact Factor, Number of Authors, and 
Type of Article. 
For doing that we construct a factorial plane by using  
the active variables (i.e. Journal Impact Factor, Number 
of Authors, and Type of Article), project onto them the 
supplementary variables (i.e. Citation Parameter 
and Topic) and interpret the results.
We performed MCA through \texttt{FactoMineR} package \citep{FactoMineR} version 2.7 in \texttt{R} version 4.1.2 \citep{Rcoreteam2021}.

Firstly, we categorized the variables 
(Table \ref{table_mca_categorization}).
The cutoffs are based on the quarterlies for
Journal Impact Factor and Citation Parameter.  
Concerning Number of Authors, the first cutoff divides 
papers with a single author, since they are the most 
desirable publication in most areas of academia 
\cite{thatje2016reaching}, 
whilst the second is the median.

\begin{table}[!ht]
    \centering
    \begin{tabular}{|l|l|l|}
    \hline
        \textbf{Active Variables} & \textbf{Modalities} & \textbf{Range} \\ \hline
        
        ~ & LowJIF & [min,1] \\ 
        Journal Impact Factor & MediumJIF & (1, 4] \\ 
        (Jif\_class) & HighJIF & (4, 6] \\ 
        ~ & MaxJIF & (6, Max] \\ 
        \hline
        ~ & LowAuth & [min,1] \\ 
        Number of Authors & MediumAuth & (1, 6] \\ 
        (author\_class) & HighAuth & (6, Max] \\ \hline
        Type of Article & Article &  \\ 
        (DT) & Review & \\ \hline\hline
        \textbf{Supplementary Variables} & \textbf{Modalities} & \textbf{Range} \\ \hline
        ~ & LowCP & [min, 0.21] \\ 
        Citation Parameter & MediumCP & (0.21, 0.47] \\ 
        (Cit\_rate\_class) & HighCP & (0.47, 1.08] \\ 
        ~ & MaxCP & (1.08, Max] \\ 
        \hline
        Topic & 21 topics & ~ \\ 
         \hline
    \end{tabular}
    \caption{Active and supplementary variables used for Multiple Correspondence Analysis.}
\label{table_mca_categorization}
\end{table}

Table \ref{tableInertia} shows how much inertia is explained 
by each of the six dimensions extracted from the active 
variables. We decided then to use for our analysis the 
first two dimensions, which explain 38.3\% of the total 
inertia. Table \ref{tableActiveContributions} shows the contribution of the 
active variables to the dimensions considered. 
Type of Article has almost no contribution in 
dimension 1, whilst it has the highest contribution 
for the second one. 
On the other hand, Journal Impact Factor and Number 
of Authors have a high relative contribution to both 
dimensions.

\begin{figure}[h]
\centering\includegraphics[scale=0.6]{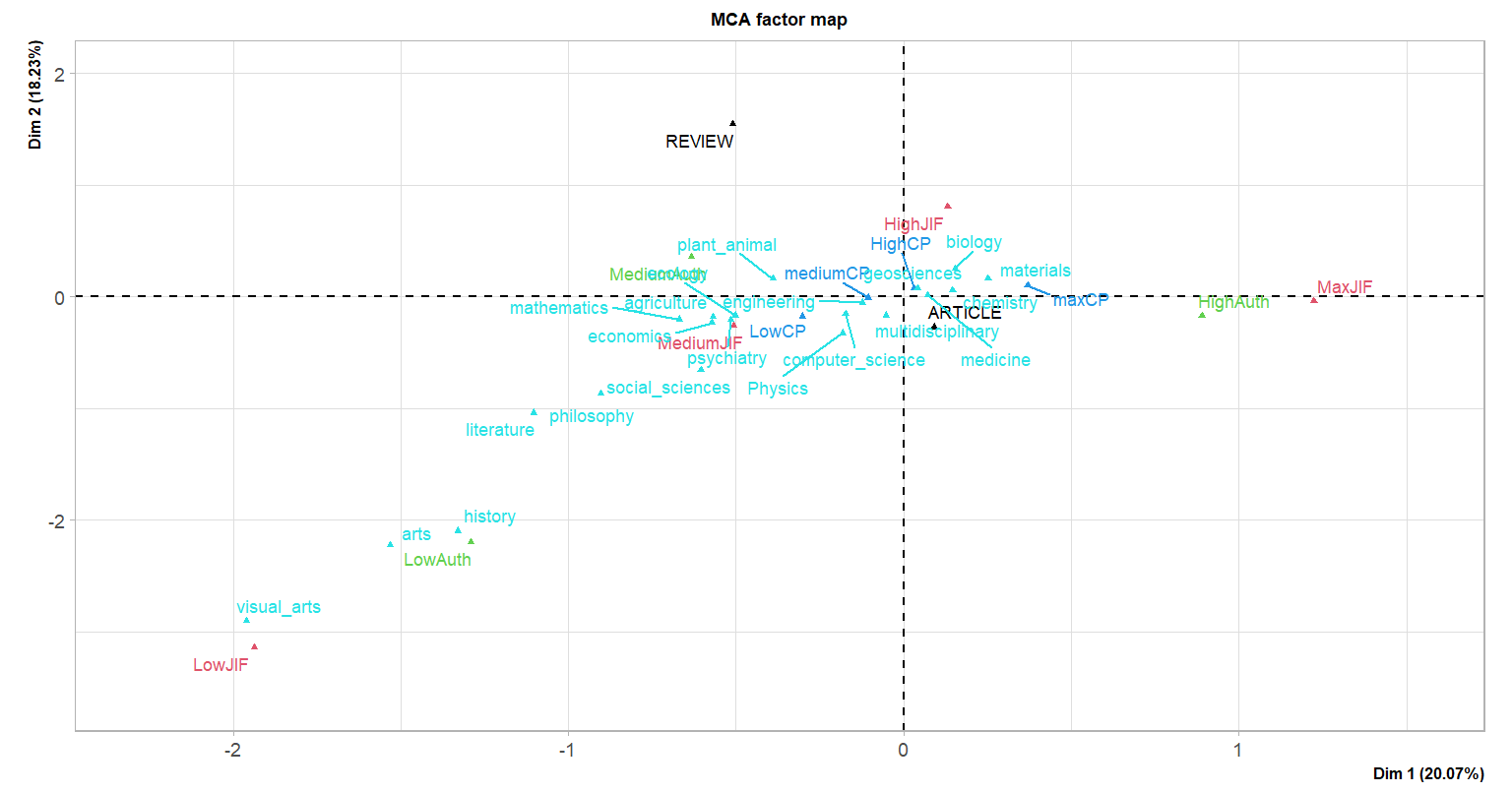}
\caption{Multiple Correspondence Analysis Bi-plot.}
\label{figureMCA}
\end{figure}

Figure \ref{figureMCA} shows the bi-plot of the dimensions
analyzed. The active variables Journal Impact Factor and 
Number of Authors appear strongly associated since the 
sorted categories of the latter are in between the sorted 
categories of the former, that is \textit{LowAuth} is in 
between \textit{LowJIF} and \textit{MediumIF}, and so on. 
Consequently, scientific publications in journals with
high impact factors tend to be signed by a large number 
of authors and vice versa, this is consistent with the
previous literature on the topic
\citep{thatje2016reaching, uthman2013citation}.
\\
The first dimension can be interpreted as the complexity 
degree of the publication, in terms of the necessary work 
behind the publication both for organizing the 
contributions of many authors and for structuring 
the work for a journal with a high impact factor.
Concerning the second dimension, instead, the main
information is provided by Type of Article, in particular
by the \textit{Review} category. High values of this
dimension correspond to review-type publications in journals with high impact factor, whilst its lowest values concern articles published in low-impact factor journals by just one author.\\
Onto this plane, we projected the two supplementary 
variables (Citation parameter, and Topics) to investigate
their association with the active variables.\\
As the first dimension is characterized by the complexity 
degree of the publication, the cloud of the modalities 
of the variable Topic shows how as we move from the
negative semi-plane to the positive one, we move from
journal topics (Arts, History, Literature) that usually 
have a low number of authors and published in low impact 
factor journals to topics such as Biology, Medicine or 
Chemistry, that are known to have articles with a large 
number of authors and refer to journals with high impact 
factors \cite{abramo2015relationship}.\\
Finally, the Citation Parameter modalities are mostly spread near the axes' origin, however, we note that the highest citation parameters are associated more with a high number of authors and high-impact factor journals.

\subsection{Analysis of citations}

We consider a regressive model to investigate the behaviour of the rate of citations per month of each paper with respect to the number of authors, the bibliometric impact of the Journal of publication in terms of Impact Factor and the different JCR research groups. 

The data collected was mainly related to Medical and Biological domains since over a total of 98,054 articles the 77\% was associated to them.
Since the beginning of the pandemic it was not hard to guess that the spread of the COVID-19 would have had an impact in medical and related research fields. However it was less predictable how research in other fields would have been also highly affected by this phenomenon. With the aim of understanding the citation dynamics in the non-medical fields, we have removed those articles from the dataset. We have also decided to not consider articles classified in the more general Multidisciplinary group when no other group of research was specified. Articles classified in research domains that usually do not have a bibliometric nature and for which there were not enough observations, such as history, literature and philosophy have also been removed. The final dataset consisted in 21,848 articles.

Three separate linear quantile regression models have been estimated for the three quartiles ($\tau=\{0.25, 0.50, 0.75\}$). Standard errors were estimated bymeans of 100 bootstrap replicates. The application of such models allows to investigate the citation behaviours across its quartiles, by distinguishing the parameter estimates for highly cited papers from those corresponding to the less cited ones. Furthermore the choice of such kind of models is suitable for dealing with the asymmetry of the response, in fact we have observed that its values range from a minimum of 0.034 to a maximum of 93 with a median of 0.47 but a mean of $0.94$.

The analysis of the number of citations, as it is well known, is useful to study the dynamics of diffusion of research in scientific communities. From our analysis it is confirmed that, also when focusing on the literature related to Sars-COV2 pandemic, there is a strong positive association of citations with the number of collaborators and the impact of the journal of publication. From Table \ref{tab:world_quantile} it can be seen how, in agreement with the literature in the field (see for instance \citep{abramo2015relationship}), the impact of the number of authors tends to become negligible above certain levels of citations, i.e. the effect of this variable appeared to be not statistically significant when considering articles in the third quartile of the response distribution.  

\begin{figure}[H]
\centering
  \includegraphics[width=0.9\textwidth]{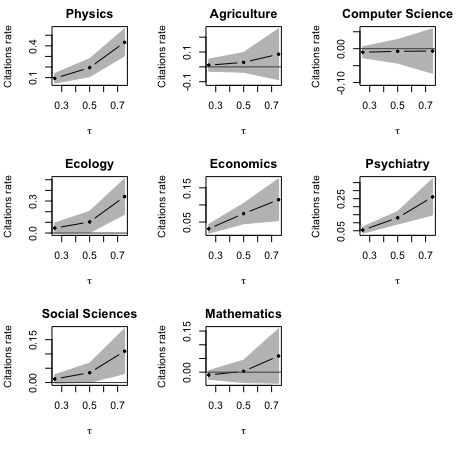}
\caption{Quartile estimates for JCR research groups. Positive trends.}
\label{Positive}
\end{figure}

\begin{figure}[H]
\centering
  \includegraphics[width=0.8\textwidth]{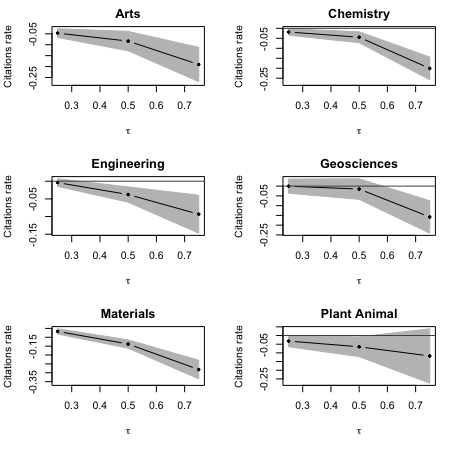}
\caption{Quartile estimates for JCR research groups. Negative trends.}
\label{Negative}
\end{figure}

What is even more interesting is that, once controlled for the effect of the number of authors and the JIF, there is a different distribution of the citations based on the JCR research groups, which also evolves across its quartiles. In particular, it emerges that, the fact of being classified in a specific discipline area can affect the rate of citations over time. In particular we observe that there are areas for which the estimated effect is positive and there is also a natural increasing trend with higher quartiles (see Figure \ref{Positive} and Table \ref{tab:world_quantile}), among them Physics, Ecology, Economics, Psychiatry and Social Sciences are the most influential. As opposite to them there are areas such as Arts, Chemistry, Engineering and Materials (see Figure \ref{Negative}), for which we observe a decrease in the citation indicator. Furthermore we can observe also differences in the magnitude of estimates.

A similar analysis was run for the Italian case study. 
After having removed the share of papers related to the fields of Biology and Medicine, those from the Multidiscipinary group and those from fields with less than 20 observations, we remain with a dataset of 765 articles. Again, a linear quantile regression for the three quartiles was estimated with these data. What was found is that some of the effects are changed with respect to what observed in the overall data set, both in sign but also in magnitude. For instance the effect of the number of authors on the rate of citations results to be negligible even for articles with few citations. What remains similar is the positive significant impact of the Physics group for articles with a median rate of citations, as well as that the negative one in the Materials group, with the same decreasing trends with the increasing of citations. It is interesting to observe also how in the Computer Science community the literature on COVID-19 seems to be negatively associated to the increase of citations. In the overall data set we observed similar estimates in terms of magnitude but they were not statistically significant. 
The results of the estimates obtained are reported in the Appendix in Table \ref{tab:italy_quantile}.

\begin{table}[H]
\begin{tabular}{|l|l|l|l|}
                   & $\tau=0.25$ & $\tau=0.50$ & $\tau=0.75$ \\
                   \hline
                   & Estimate (SD) & Estimate (SD) & Estimate (SD) \\
                   \hline
(Intercept)	& 0.090(0.011)*** &	0.164(0.020)***	& 0.347(0.041)*** \\
JIF &	0.029(0.002)***	& 0.066(0.004)*** & 0.143(0.006)*** \\
N. authors & 0.003(0.001)*	& 0.004(0.002)** &	0.008(0.005) \\
Physics &	0.093(0.031)**	& 0.194(0.055)***	& 0.433(0.074)*** \\
Agriculture & 0.011(0.022) &	0.030(0.039) & 0.085(0.103) \\
Ecology &	0.046(0.030) &	0.104(0.058). & 0.342(0.100)*** \\
Economics &	0.030(0.009)***	& 0.075(0.016)***	& 0.115(0.034)*** \\
Psychiatry &	0.055(0.014)***	& 0.131(0.024)*** & 0.261(0.064) \\
Social Sciences & 0.012(0.009) & 0.034(0.020). &	0.109(0.054)* \\
Mathematics & -0.010(0.010) &	0.003(0.025) & 0.059(0.060) \\
Geosciences & -0.0004(0.022) &	-0.015(0.030) &	-0.158(0.050)** \\
Engineering & -0.004(0.007) &	-0.038(0.012)** & -0.093(0.032)** \\
Computer Science &	-0.010(0.010) &	-0.008(0.021) &	-0.007(0.039) \\
Arts &	-0.047(0.013)*** &	-0.083(0.028)** & -0.190(0.045)*** \\
Chemistry &	-0.018(0.009)* &	-0.044(0.017)** & -0.201(0.037)*** \\
Materials & -0.067(0.009)*** &	-0.138(0.016)*** & -0.282(0.033)***\\
Plant Animal &	-0.032(0.018). &	-0.065(0.037). & -0.118(0.093) \\
\hline
\end{tabular}
\caption{Quantile regression model estimates}
\label{tab:world_quantile}
\end{table}

\section{Discussion and conclusions}
\noindent 
In this paper we have inspected the impact of the outbreak of the COVID-19 pandemic in the global scientific literature available from the bibliometric archive of Web of Science under different points of view. We presented an updated picture of the COVID-19 literature both at global and local level by conducting a specific case study on Italy, which was highly affected in the early phases of the pandemic. As expected, the research domains more affected by this phenomenon were associated to the medical fields, the epidemiology and health in general. 
Globally, the scientific production is led by the US and China (in accordance with previous literature). We observed how the collaboration network among countries is widespread all over the world. However, we also showed geographically distinct clusters of collaboration among universities or other institutions. The red cluster only included institutions based in USA, while the blue cluster included institutions from different countries (e.g. United Kingdom, Canada, Italy, Brazil, India and Hong Kong). The other two smaller clusters only contained institutions from Asian countries (green and purple clusters). We can therefore conclude that countries such as e.g. US or China showed stronger internal collaborations compared to countries such as Canada, UK or Germany, as also indicated by the higher multiple country collaboration ratio obtained by the latter countries. 

Finally, using MCA and quantile regression, we evaluated the relationship between the number of citations and other variables (mainly the journal impact factor, the number of authors, and the topics of the journal). Using MCA, we observed the highest citation parameters to be associated with a high number of authors and high-impact factor journals. In addition, we showed that the modalities of the variable Topic ranged from topics that usually have a low number of authors and are published in low impact factor journals (such as Arts, History, Literature) to topics often characterized by articles with a large number of authors and published in journals with high impact factor (such as Biology, Medicine or Chemistry). Quantile regression analysis confirmed the strong relationship of citations with the number of authors and the JIF in non-medical fields, but highlighted that above a certain threshold, i.e. for highly cited articles, the impact of the number of authors almost becomes irrelevant. The analysis showed also how the spread of citations on the topic varies according to the research domain with distinct magnitudes across the quartiles of the response.  

This work presents some limitations. The exact publication date of some articles (e.g. articles published with an early publication procedure) might be different from the one available in WoS (which only reports the publication date of the issue). This might result in an underestimation of the lifespan of a scientific article and therefore a higher parameterized number of citations. Moreover, as regards to the analyses concerning the data set of authors with an Italian affiliation, it was not possible to obtain the SSD for all authors as some do not work in universities (e.g. doctors, engineers) or their names/last names were misspelled or abbreviated.

\clearpage
\bibliography{biblio_covid19}

\section{Appendix}

\begin{table}[ht]
\centering
\begin{tabular}{rrrr}
  \hline
 & Eigenvalue & \% of variance & Cumulative \% of variance \\ 
  \hline
dim 1 & 0.40 & 20.07 & 20.07 \\ 
  dim 2 & 0.36 & 18.23 & 38.30 \\ 
  dim 3 & 0.35 & 17.30 & 55.59 \\ 
  dim 4 & 0.32 & 16.20 & 71.80 \\ 
  dim 5 & 0.31 & 15.55 & 87.35 \\ 
  dim 6 & 0.25 & 12.65 & 100.00 \\ 
   \hline
\end{tabular}
\caption{Inertia explanation of the six dimensions.}
\label{tableInertia}
\end{table}

\begin{table}[ht]
\centering
\begin{tabular}{rrr}
  \hline
 & Dim 1 & Dim 2 \\ 
  \hline
Type of Article & 0.05 & 0.42 \\ 
  Journal Impact Factor & 0.52 & 0.35 \\ 
  Number of Authors & 0.63 & 0.32 \\ 
   \hline
\end{tabular}
\caption{Active variables contributions to the first two dimensions.}
\label{tableActiveContributions}
\end{table}

\begin{table}[h]
\begin{tabular}{|l|l|l|l|}
                   & $\tau=0.25$ & $\tau=0.50$ & $\tau=0.75$ \\
                   \hline
                   & Estimate (SD) & Estimate (SD) & Estimate (SD) \\
                   \hline
(Intercept)	& 0.121(0.048)*	& 0.265 (0.069)*** & 0.502 (0.195)* \\
JIF &	0.015 (0.008). & 0.043 (0.010)*** &	0.076 (0.027)**  \\
N. authors & 0 (0.001) & -0.001 (0.004) & 0.001 (0.009)\\
Physics &	0.136 (0.131) & 0.260 (0.119)* & 0.358 (0.233) \\
Economics &	0 (0.030) &	-0.053 (0.072) & -0.029 (0.149) \\
Psychiatry & 0.036 (0.053) & -0.002 (0.115) & 0.306 (0.245) \\
Social Sciences & 0.040 (0.048)	& 0.012 (0.081)	& 0.005 (0.214) \\
Mathematics & 0.002 (0.037) & -0.049 (0.085) & -0.026 (0.177) \\
Engineering & 0.007 (0.019)	& -0.059 (0.055) & 0.060 (0.122) \\
Computer Science &	-0.063 (0.031)* & -0.161 (0.077)* &	-0.199 (0.134) \\
Chemistry &	0.001 (0.029) &	-0.066 (0.057) & -0.095 (0.130) \\
Materials & -0.067 (0.029)* & -0.165 (0.066)* & -0.342 (0.128)**\\
\hline
\end{tabular}
\caption{Quantile regression model estimates for the subgroup of articles coauthored by at least one Italian researcher}
\label{tab:italy_quantile}
\end{table}

\end{document}